\documentclass{ws-ijmpd}
\usepackage[super,compress]{cite}
\usepackage{amssymb}
\usepackage{graphicx}	
\usepackage{amsmath}
\usepackage{color}

\def\comment#1{}

\begin{document}

\markboth{W.-B. Han, S.-C. Yang}
{Exotic orbits due to spin-spin coupling around  Kerr black holes}

%
\catchline{}{}{}{}{}
%

\title{Exotic orbits due to spin-spin coupling around \\ Kerr black holes }

\author{Wen-Biao Han* and Shu-Cheng Yang}

\address{Shanghai Astronomical Observatory,\\ 
CAS, Shanghai, 200030, P. R. China\\
School of Astronomy and Space Science,\\
University of Chinese Academy of Sciences,\\
Beijing 100049, P. R. China\\
*wbhan@shao.ac.cn}

\maketitle

\begin{history}
\received{8 June 2017}
\revised{14 August 2017}
\accepted{17 August 2017}
Published
\end{history}

\begin{abstract}
We report exotic orbital phenomena of spinning test particles orbiting around a Kerr black hole, i.e., some orbits of spinning particles are asymmetrical about the equatorial plane. When a nonspinning test particle orbits around a Kerr black hole in a strong field region, due to relativistic orbital precessions, the pattern of trajectories is symmetrical about the equatorial plane of the Kerr black hole. However, the patterns of the spinning particles' orbit are no longer symmetrical about the equatorial plane for some orbital configurations and large spins. We argue that these asymmetrical patterns come from the spin-spin interactions between spinning particles and Kerr black holes, because the directions of spin-spin forces can be arbitrary, and distribute asymmetrically about the equatorial plane. 
\end{abstract}


\ccode{PACS number(s): 04.20.Cu, 04.25.dg, 04.25.dk, 97.80.-d}


\section{Introduction}
The equations of motion of a nonspinning test particle around a Kerr black hole are fully integrable because of the existence of four conserved quantities: rest mass, energy, angular momentum and the Carter constant \cite{carter68}.  The axisymmetry of spacetime drives the geodesic orbits to fill the volume in an axisymmetric manner. The same holds for the reflection symmetry of the background along the equatorial plane. As a result, in the strong gravitational field region, due to two relativistic orbital precessions: perihelion and Lense-Thirring precessions which reflect the spacetime symmetries, the pattern of trajectories of the test particle is symmetrical about both the rotating axis and equatorial plane of the corresponding Kerr black hole, i.e., after many laps the Kerr geodesic orbits crudely fill a volume that is loosely symmetric about the polar axis and equatorial plane. In principle, even for the zone far from black holes, providing a sufficient time scale, all orbital configurations of test particles around Kerr black holes have these two symmetries. 

Though almost all astrophysical bodies have spins, in the region far away from the central massive body, for extreme mass-ratio cases, the motion of a small body can be described accurately enough with the nonspinning test particle approximation. For example, the spins of S-stars around the supermassive black hole in our Galactic center \cite{shen05,ghez08,gillessent09,genzel10,han14} can be ignored. 

However, in the strong field region and a large spin, due to the spin-orbit and spin-spin interactions, the trajectories of a spinning particle in Kerr spacetime can deviate from geodesic motion. Unlike the nonspinning case, for the spinning particles, because of the extra degrees of freedom caused by the spin vector and absence of the Carter constant, the equations of motion of spinning particles are no longer integrable. The spin of the particle is important in dynamics and gravitational waves for extreme mass-ratio systems \cite{tanaka96,bini04,han10,huerta11,hackmann14,hughes15,harms16}. For the nonspinning case,  the orbits around a Kerr black hole are always regular. However, under some conditions, and for extreme spin values the orbital motions of extreme spinning particles can be chaotic (see Refs. 14--19 and references inside). Such extreme spin values are actually impossible for compact objects like black holes, neutron stars, white dwarfs etc. For noncompact bodies like planets, the spin magnitude can approach 1 (in our units,   see next paragraph), for example, the Jupiter-Sun system. Unfortunately, due to the tidal influence from the central black hole, such a noncompact body will be disrupted by the black hole in the strong field region (see Sec. 2 for details). Therefore, for relativistic large-mass-ratio binary systems, the spin magnitude of the small object should be much less than $1$.  However, the phenomena and characteristics of extreme spinning particles orbiting near a black hole are very interesting for researchers \cite{suzuki97,suzuki98,hartl03,hartl04,han08,georgios16,semerak99,semerak07}. In this paper, we focus on an exotic orbital configuration whose orbit pattern is asymmetrical about the equatorial plane of the Kerr black hole.  We try to study this interesting phenomenon in details and reveal its physical reasons. 

Through this paper, we use units where $G = c = 1$ and sign conventions $(-, +, +,+)$. The time and space scale is measured by the mass of black hole $M$, and energy of particle is measured by it's mass $\mu$, the angular momentum and spin by $\mu M$, and linear momentum by $\mu$. We also assume that $\mu / M \ll 1$.

\section{Mathisson-Papapetrou-Dixon equations and repulsive effect from spin-spin coupling}
The popular equations for describing the motion of a spinning particle in curved space-time are Mathisson-Papapetrou-Dixon (MPD) equations \cite{mpd1,mpd2,mpd3,mpd4},
\begin{align} 
\frac{D p^\mu} {D \tau} &= -\frac{1}{2} R^\mu_{~\nu\rho\sigma} \upsilon ^\nu S^{\rho\sigma} \,, \label{mpdeq1} \\
\frac{D S^{\mu\nu}}{D\tau} &= p^\mu \upsilon^\nu - \upsilon^\nu p^\mu \,, \label{mpdeq2}
\end{align}
where $\upsilon^\nu \equiv dx^\nu/d\tau$ is the four-velocity if $\tau$ is the proper time of the spinning particle, $p^\mu$ the linear momentum, $S^{\mu\nu}$ the anti-symmetrical spin tensor, and $R^\mu_{~\nu\rho\sigma}$ the Riemann tensor of the background. Alternative approaches to the spinning particle equations can be found in Refs. 26 and 27. The spin tensor is then related with the spin vector by
\begin{align}
S^{\mu\nu} =  \epsilon^{\mu\nu\alpha\beta} u_\alpha S_{\beta} \,,
\end{align}
where $u_{\alpha} \equiv p_{\alpha}/\mu$, $\epsilon^{\mu\nu\alpha\beta} = \varepsilon^{\mu\nu\alpha\beta}/\sqrt{-g}$ is a tensor and $\varepsilon^{\mu\nu\alpha\beta}$ the Levi-Civita alternating symbol ($\varepsilon_{0123} \equiv 1,~\varepsilon^{0123} \equiv -1$).  Following Tulczyjew \cite{Tulczyjew}, we choose 
\begin{align}
p^\mu S_{\mu\nu} = 0 \Longrightarrow p^\mu S_\mu = 0 \label{ps}
\end{align}
as a spin supplementary condition which defines  a unique worldline identified with the center of mass.
Condition (\ref{ps}) leads to the velocity-momentum relation (see e.g. Ref. 20)
\begin{align}
\upsilon^\mu = \frac{m}{\mu}\left(u^\mu+ \frac{2S^{\mu\nu}R_{\nu\sigma\kappa\lambda}u^\sigma S^{\kappa\lambda}}{4{\mu}+R_{\alpha\beta\gamma\delta}S^{\alpha\beta}S^{\gamma\delta}}\right) \,, \label{vp}
\end{align}
where $\mu$ is the ``dynamical" rest mass of the particle defined by $p^\nu p_\nu = -\mu^2$ and is a constant here because of the supplementary condition we chose. $m$ is the ``kinematical" mass which is not a constant and defined by $p^\nu \upsilon_\nu = -m$. In order to obtain the four-velocity through Eq. (\ref{vp}), one normalizes $m$ in such way that $v^\mu v_\mu=-1$. One can see Ref. 20 for detailed discussions. Now, the MPD equations become a closed form and can be calculated. 

Due to the lack of enough conserved quantities, there is no analytical solution for Eqs.(\ref{mpdeq1}) and (\ref{mpdeq2}), and then numerical integration is used to calculate the motion of the spinning particle.  We choose  Boyer-Lindquist coordinates for calculations, and firstly we should give a set of initial conditions at time $t_0$: $r_0, \theta_0, \phi_0, u^{\mu}_0$ and $S^{\mu}_0$. It is noted that there are three constraints and two constants of motion:  the constraints $u^\mu u_\mu = -1, S^\mu S_\mu =S^2$ (S is the spin magnitude), and the spin supplementary condition (\ref{ps}), as well as the energy and angular momentum constants, because of the two Killing vectors $\xi^\mu_t, ~\xi^\mu_\phi$. The energy and angular momentum are given as (e.g. Refs. 20 and 21),
\begin{align}
E=-p_t+\frac{1}{2} g_{t\mu,\nu}S^{\mu\nu}, \\
L_z=p_\phi-\frac{1}{2} g_{\phi\mu,\nu}S^{\mu\nu}. 
\end{align} 
Hereafter, we use the dimensionless quantities $u^\nu$ to replace $p^\nu$, because we basically can utilize the dimensionless counterparts of the quantities presented up to this point. Note that in numerics the dimensionless and the dimensionful quantities are equivalent if one sets $\mu=M=1$.

As a result, for setting the initial conditions, three components of $u^{\mu}_0$, two components of $S^{\mu}_0$ are not arbitrary, they must satisfy the above five constraint and conservation equations. In this paper, we set a group of initial conditions by hand: $r_0, \theta_0, \phi_0, u_0^\theta, S_0^r ~{\rm and} ~S_0^\theta$, and also the energy $E$, the orbital angular momentum $L_z$ and the spin magnitude $S$. From the five mentioned equations, we can solve out the left initial conditions: $u^t, u^r, u^\phi$ and $S^t, S^\phi$. The relative accuracy of the calculated initial conditions can achieve $10^{-15}$ with double precision codes. With these initial conditions, one can immediately get $\upsilon^\mu$ with the help of Eq. (\ref{vp}). In every numerical step, we integrate Eqs. (\ref{mpdeq1}) and (\ref{mpdeq2}) and solve the velocity-momentum relation (\ref{vp}) at the same time.  At the end of numerical evolution, all these conserved quantities must be checked again to make sure the calculations are accurate enough. During our numerical simulations, the relative errors of all these constraints are about $10^{-13}$ after $\tau = 10^6~M$ evolution. 

\begin{figure}
\begin{center}
\includegraphics[height=2.2in]{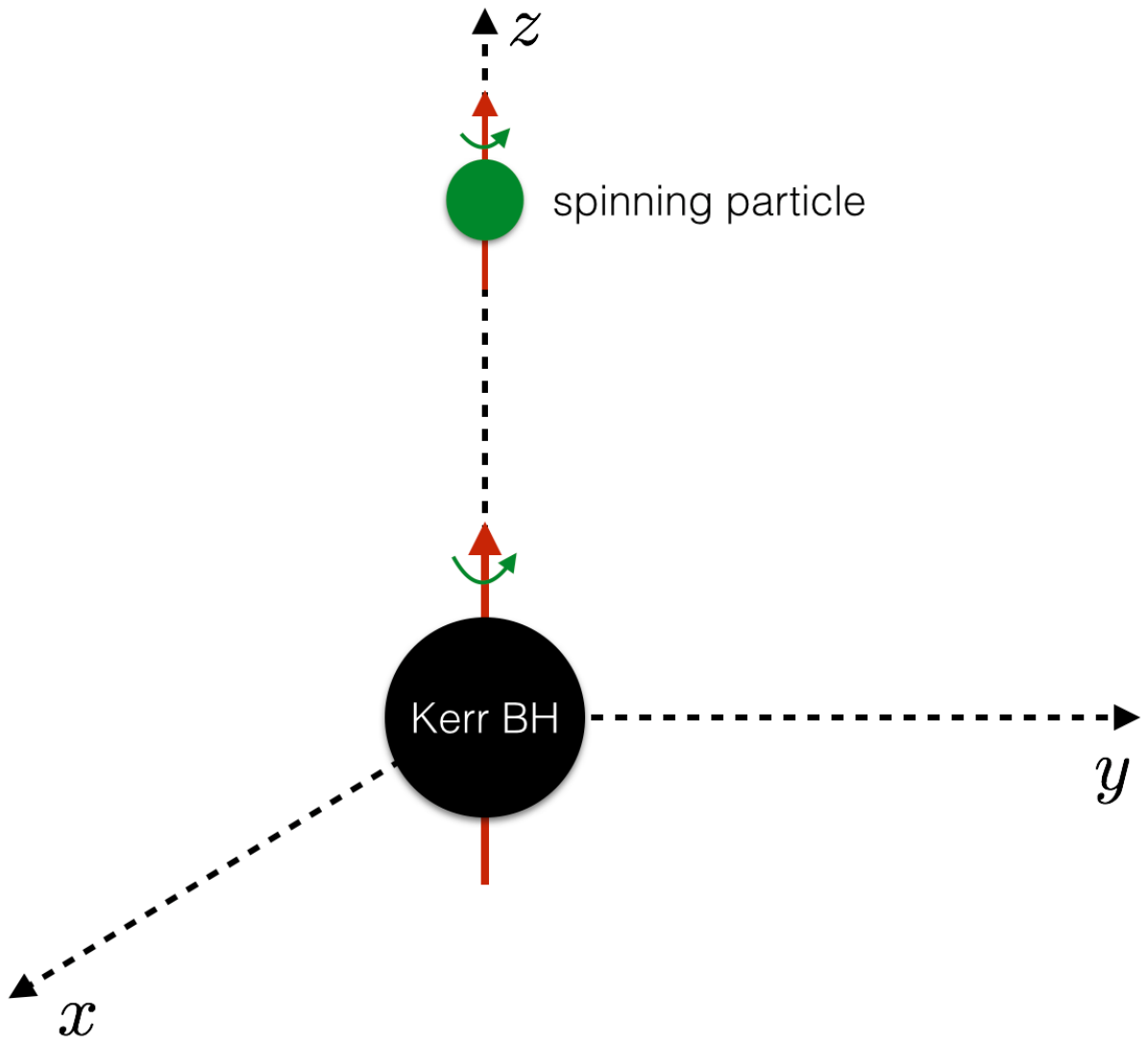}
\caption{A spinning particle with aligned spin to a Kerr black hole locates at the $z$-axis.}  \label{spinspin}
\end{center}
\end{figure}

For simplification, firstly, we assume a spinning particle locating at the polar direction of the Kerr black hole (i.e. $\theta  = 0$). Because of the bad behavior of Boyer-Lindquist coordinates at $\theta = 0$, we transfer the coordinates to Cartesian-Kerr-Schild ones, then the line element is written in $(t, x, y, z)$ as \cite{cks}
\begin{align} \nonumber
ds^2 = &-dt^2 + dx^2 + dy^2 + dz^2  \\ 
           &+\frac{2Mr^2}{r^4+a^2z^2}  \left[dt+\frac{r(xdx+ydy)}{a^2+r^2}+\frac{a(ydx-xdy)}{a^2+r^2}+\frac{z}{r}dz\right]^2.
\end{align}
In the Cartesian coordinates, the spinning particle is put at $(0,~0,~z_0)$ originally, the components of the initial $u^\mu$ are all zero except for $u^t$, and the only nonzero component of the spin vector is $S^z$. From Eq. (\ref{vp}), the only nonzero component of the four-velocity is $\upsilon^t$. The schematic diagram of an aligned spin configuration is shown in Fig. \ref{spinspin}. 

Based on these assumptions, Eq. (\ref{mpdeq1}) is reduced to 
\begin{align}
\frac{d u^z} {d\tau} &= -u^t\upsilon^t\left(\Gamma^z_{tt}+\frac{1}{2}SR^z_{~txy}  \frac{g_{tt}}{\sqrt{g^{zz}}}\right) \equiv u^t\upsilon^t (F_{\rm m}+F_{\rm ss})\,,  
\end{align}
where $F_{\rm m}$ means the gravitational interaction due to the curvature (mass) and $F_{\rm ss}$ the spin-spin interaction. The second term is definitely zero when $S=0$ or $a=0$. For clarity, we write down the expressions of them
\begin{align}
F_{\rm m} &= - M\frac{(z^2-a^2)(z^2-2 M z+a^2)}{(z^2+a^2)^3} \, , \\
F_{\rm ss} &= +MaS\sqrt{\frac{z^2-2Mz+a^2}{z^2+a^2}}\frac{[z^3(3z-6M)+2a^2z(z+M)]-a^4}{(z^2+a^2)^4}\label{eleven}\, .
\end{align}

\begin{figure}
\begin{center}
\includegraphics[height=1.4in]{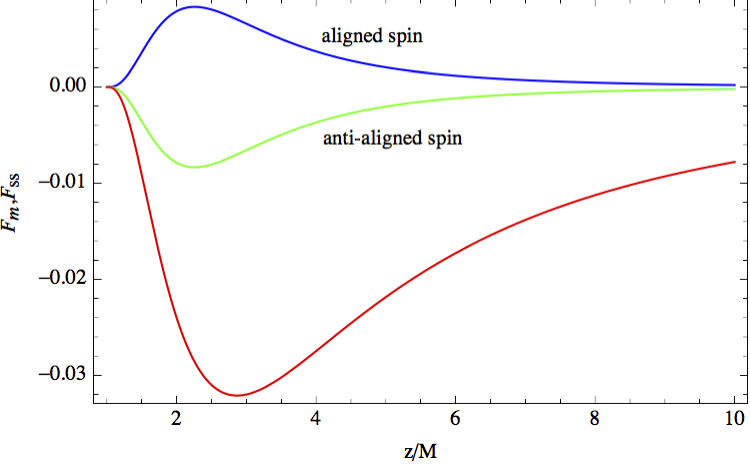}
\includegraphics[height=1.4in]{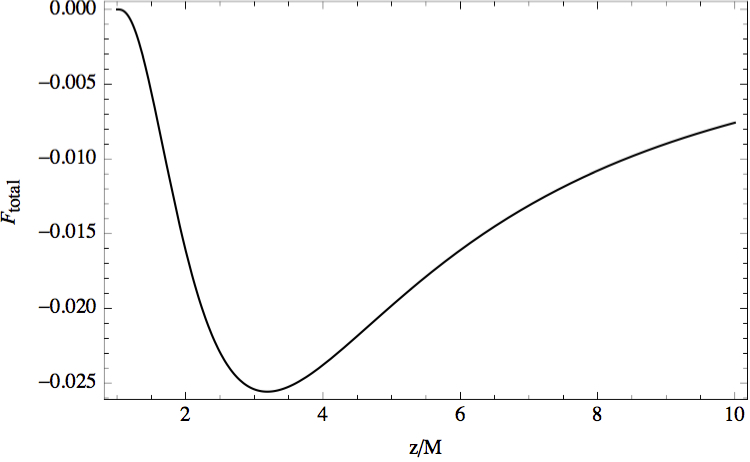}
\caption{(Color online) Left panel: $F_{\rm m}$ (red solid line), $F_{\rm ss}$ (aligned and anti-aligned cases); Right panel: $F_{\rm m}+F_{\rm ss}$ for the aligned spin case (directions of the spins of particle and black hole are the same). The parameters used for plotting are $a=1, |S|=1$.}  \label{forces}
\end{center}
\end{figure}

The behaviors of these two functions near horizon are plotted in Fig. \ref{forces}. Outside of the horizon, the value of $F_{\rm m}$ is always negative to offer ``regular" gravity (here we assume $z$ is positive, for negative $z$, vice versa). However, we can clearly find that the spin-spin coupling force $F_{\rm ss}$ is positive with aligned spin. If we change the direction of spin, the direction of spin-spin coupling is also changed (See the anti-aligned case in Fig. \ref{forces}). In this way, the spin-spin coupling can be thought as a kind of phenomenological counter-gravity. However, when the spin value $\leq 1$, the spin-spin force  is not as large as the interaction induced by the mass so that it cannot fully counteract the latter one. 

Actually, the physically allowed value of spin of the particle in the extreme-mass-ratio system should be much less than 1. For compact objects like black holes, neutron stars or white dwarfs, the magnitudes of spins are $\sim \mu^2/\mu M = \mu/M \ll 1$ \cite{hartl03}. However, for a noncompact body like Jupiter, the spin value of it in the Jupiter-stellar mass black hole system can be as large as 1. For this case, in the ultra-relativistic region, the tidal influence from the black hole cannot be ignored. The tidal radius $r_t \sim R_{\rm p} (M/\mu)^{1/3} \gg R_{\rm s}$ ($R_{\rm p}$ is the radius of planet and $R_{\rm s}$ the Schwarzschild radius of the black hole, see Eq. (6.1) in Ref. 30), then the planet will be disrupted by the black hole in the strong field region.

Even though the astrophysically relevant dimensionless values of the spin should be much less than 1, in the study we are interested in the dynamical aspects of the MPD equations and these aspects $S\sim 1$. For example, in Fig. \ref{forces}, for the spin magnitude $S=1$ and an extreme Kerr black hole, we find  that $F_{\rm ss}$ is always less than $F_{\rm m}$. Mathematically, equilibrium points between the two forces do not exist until the spin reaches the value 2.4925. This spin value is impossible for an extreme-mass-ratio system involving stellar compact objects, so there is no equilibrium point for such systems. On the other hand, there is no equilibrium point for noncompact objects, too. As analyzed by Wald, the MPD equations will be invalid in this extremely large spin cases, because  this spin magnitude asks for a body whose size is greater than the back ground curvature (see Ref. 31).

Generally, the direction of the spin-spin coupling can be arbitrary due to the orientation of the spin vector, unlike the mass part, which always points to the mass center. Along the z axis, only the spin-spin interaction is present; a spin-orbit interaction will appear if the small body is moved off the axis. It should be mentioned here that several papers \cite{wald72,tanaka96,costa16} have already discussed this situation, and the results in this paper coincide with theirs.

Now, we analyze the ``acceleration" along the $\theta$ direction ($du^\theta/d\tau$) when a spinning particles lies on the equatorial plane. For convenience, we are back to the Boyer-Lindquist coordinates. The contribution of the curvature part on $du^\theta/d\tau$ is $-u^\theta v^r/r$, which is reflectively symmetrical about the equatorial plane. When the sign of $u^\theta$ is changed , the sign of $du^\theta/d\tau$ is also changed but the magnitude remains unchanged. 

For simplification, we set the initial $v^r = 0$, then the contribution of the curvature part disappears. If we allow the spin direction of particle to be arbitrary (no longer aligned with the rotational axis of Kerr black hole), the contribution from spin-curvature coupling on $du^\theta/d\tau$ (i.e. the right hand of Eq. (\ref{mpdeq1})) is nonzero:
\begin{align}
 \nonumber
\frac{du^\theta}{d\tau} = & \frac{MS_r}{r^7} \{v^\phi[(L_z-aE)(3a^3-2aMr)+ar^2(3L_z-5aE)-2r^4E] \\ 
&-v^t[(L_z-aE)(3a^2-2Mr)+r^2(L_z-3aE)] \} \,. \label{twelve}
\end{align}
We notice that the above equation does not include $u^\theta$. In the first-order approximation of S, $v^{t,\phi}\approx+u^{t,\phi}+O(S^2)$, Eq. (\ref{twelve}) is independent from $u^\theta$. Actually, under our assumption $v^r = 0, \theta =\pi/2$, the right-hand side of (\ref{twelve}) is independent from $u^\theta$  exactly, though the mathematical proof is complicated (solve $v^t$, $v^\phi$ from Eq.(\ref{vp}) and take into Eq. (\ref{twelve}). We will see this point in the following numerical experiments. This means when $u^\theta$ changes sign, the ``acceleration" contributed by the spin $du^\theta/ d\tau  (s)$ does not change its sign and at the same time the magnitude remains. In this sense, the reflection symmetry is destroyed due to the spin of the particle. 
\begin{figure}
\begin{center}
\vspace{-33mm}
\includegraphics[height=4.0in]{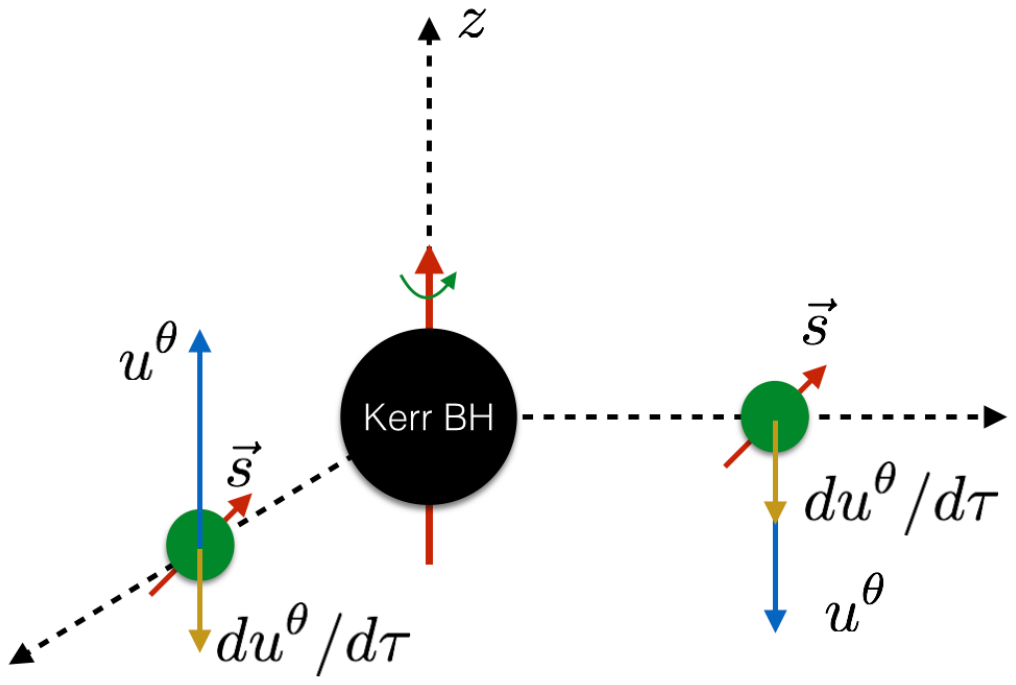}
\vspace{-30mm}
\caption{Spinning particles locates at the equatorial plane of a Kerr black hole.}  \label{equator}
\end{center}
\end{figure}

One can see an example demonstrated in Fig. \ref{equator}. The direction angle of spin is fixed as $\hat{\alpha}^s = 83.1^\circ, ~|\hat{\beta}^s| =52.9^\circ$ (see the next section for the details of our definition of spin direction), and $u^r = v^r =0$ at this moment. When $u^\theta$ changes its value from $3.8\times10^{-2}$ to $-3.8\times10^{-2}$, $ du^\theta/d\tau$ does not change its sign (always equals $ 3.0\times10^{-5}$). This property may give a clue to the exotic orbits studied in this paper.\\

\section{Exotically asymmetrical orbits}
\begin{figure}
\begin{center}
\vspace{-20mm}
\includegraphics[height=3.0in]{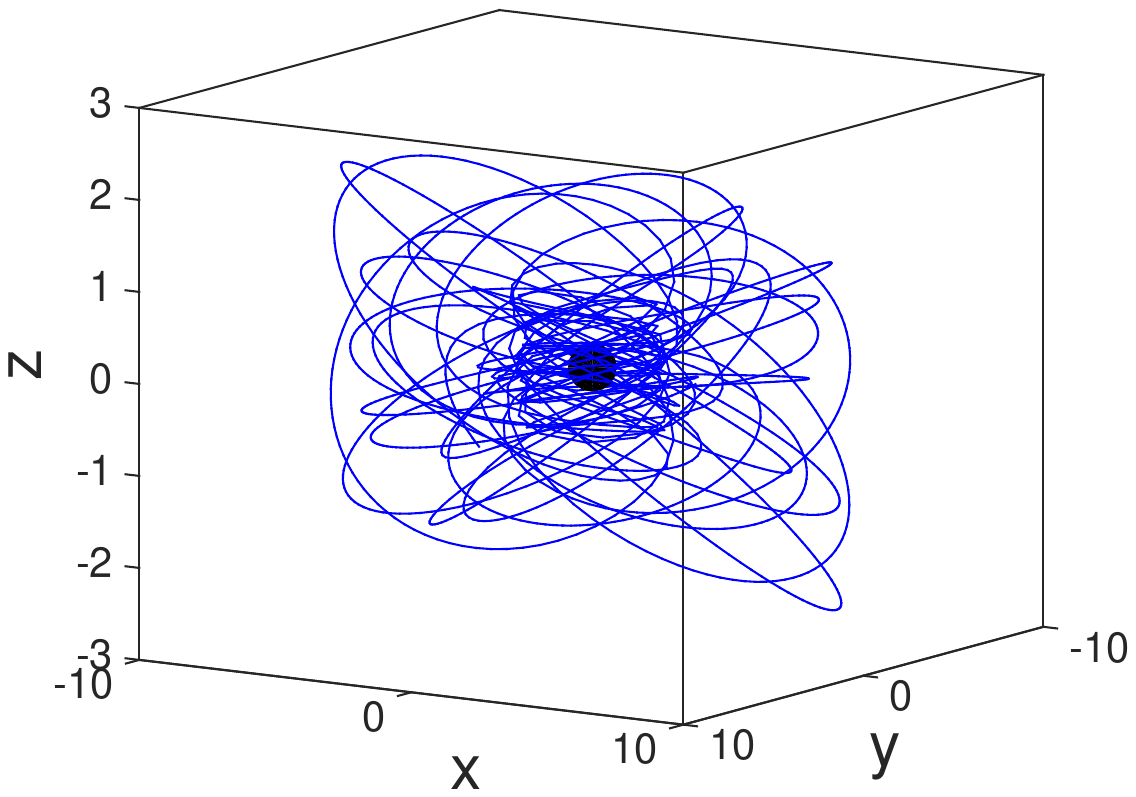}
\vspace{-30mm}
\includegraphics[height=3.0in]{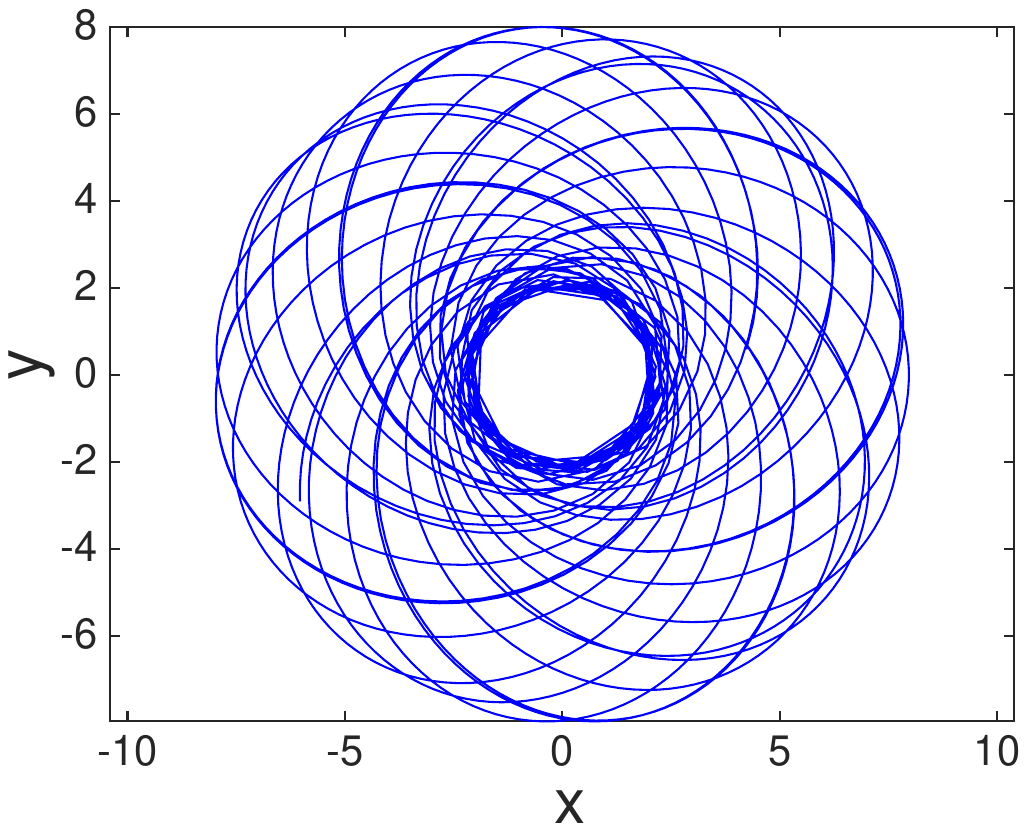}
\vspace{-10mm}
\includegraphics[height=3.0in]{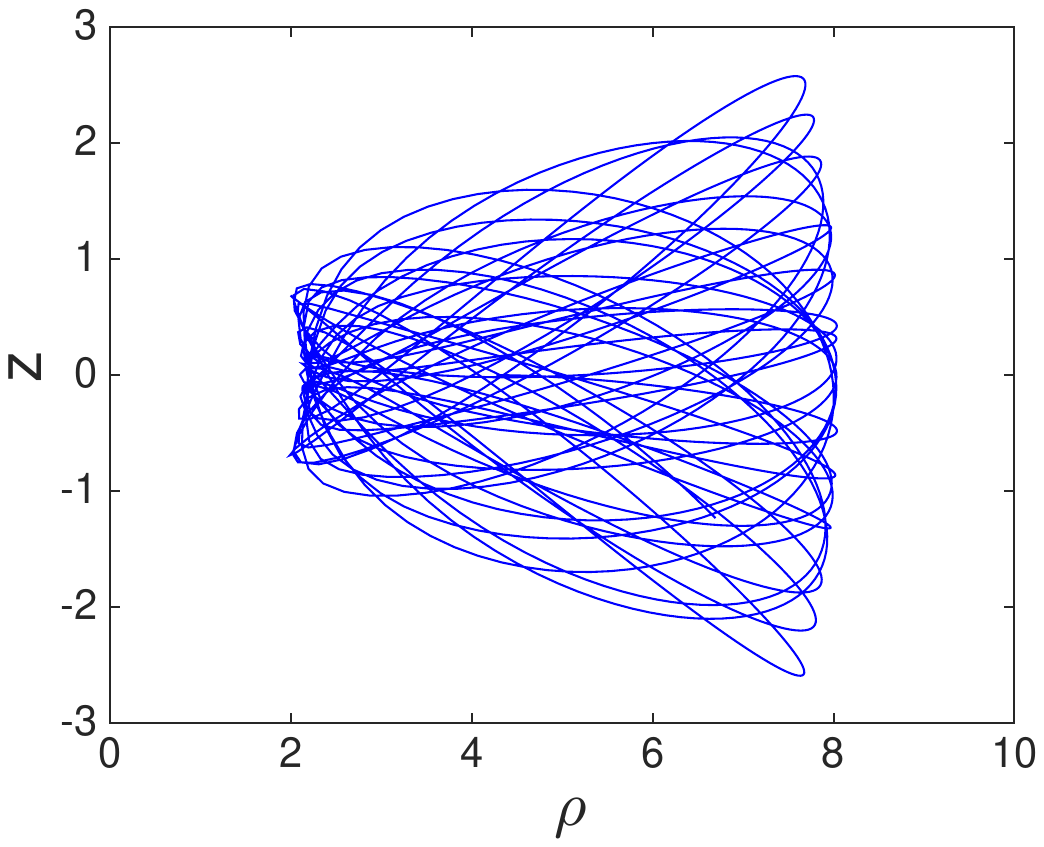}
\includegraphics[height=3.0in]{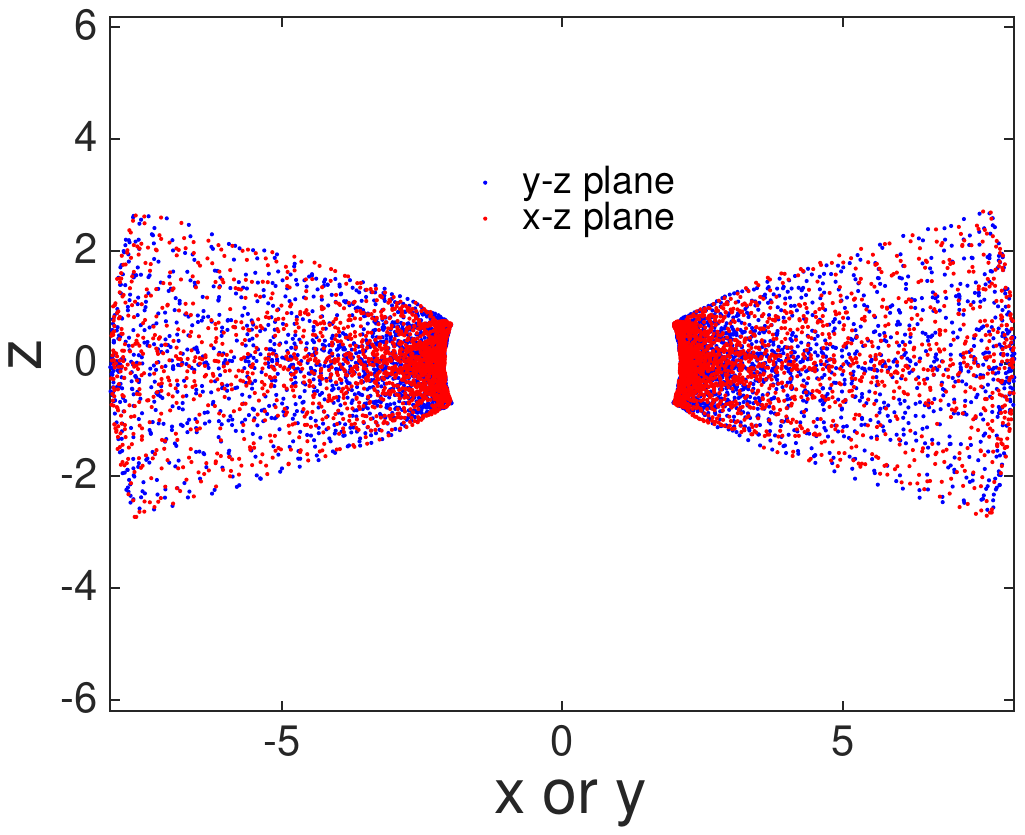}
\caption{Orbits of a spinning particle with spin parameter $S = 0.8 $, energy $E = 0.9$ and total angular momentum $L_z = 2.5$ around a Kerr black hole with $a = 1$. The initial conditions are $r_0 = 4,~\theta_0=\pi/2, \phi_0 =0$, $u_0^\theta = 0$ and $S_0^{r,\theta} =  (-0.32, ~-0.16)$. The subsidiary data are 1.623, 0.285, 0.156, -0.390 and -0.096 for $u^t_0, ~u^r$, $u^\phi$, $S^t$ and $S^\phi$, respectively. The corresponding initial angles $\hat{\alpha}^s_0, ~\hat{\beta}^s_0$ are $120.2^\circ, ~29.1^\circ$. The top-left panel shows the 3D trajectories, the top-right and bottom-left ones show the projection orbits on $x-y$ and $\rho-z$ planes respectively, where $\rho =\sqrt{x^2+y^2}$. The bottom-right panel shows the projection points when the trajectories pass through the $y-z$ and $x-z$ planes.}  \label{symspin}
\end{center}
\end{figure}

As we have mentioned before, due to the axisymmetry  of the spacetime, and  the reflection symmetry of the background along the equatorial plane, for nonspinning test particles orbiting the Kerr black hole, the perihelion advance makes the pericenter to precess in the orbital plane, and the frame-dragging effect causes the orbital plane to precess around the rotation axis of the Kerr black hole at the same time. Because of these two precessions, the patterns of the particles' trajectories distribute symmetrically about the equatorial plane and the rotation axis of the black hole.  For the spinning particles, even for the highly spinning ones, in most cases, the patterns still have these two symmetries (sometimes approximately). Figure \ref{symspin} shows the orbit of a spinning particle with $S=0.8$, energy $E = 0.9$ and total angular momentum $L_z = 2.5$ around an extreme Kerr black hole with $a = 1$, and clearly demonstrates the symmetry. 

For the numerical calculation of the orbits,  first we need to input the initial conditions. The free parameters inputted by hand are the initial coordinates $t_0=0, r_0, \theta_0, \phi_0$, one initial component of the four velocity $u^\theta_0$, two initial components of the spin vector $S^r_0, ~S^\theta_0$, and the values of $E, ~L_z$ and $S$. The remaining five initial conditions $u^t_0, u^r_0, u^\phi_0$ and $S^t_0,~S^\phi_0$ are subsidiary quantities which are calculated from the five constraint equations. We also compute the initial direction of  the spin from the initial conditions for understanding better the spin vector. For describing the direction of spin, we introduce a local hypersurface-orthogonal observer (HOO). In a Kerr space-time, the HOO is represented by an observer with zero angular momentum with respect to the symmetry axis, ZAMO, having

\begin{align}
u^\mu_{\rm ZAMO} = \sqrt{\frac{A}{\Delta\Sigma}} (\frac{1,0,0,2Mar}{A}) \,,
\end{align}
where $\Delta=r^2-2Mr+a^2$, $\Sigma=r^2+a^2\cos^2\theta$ and $A=(r^2+a^2)^2-\Delta a^2\sin^2\theta$. The relative spin with respect to the HOO is given as
\begin{align}
\hat{S}^\mu = \hat{\Gamma}^{-1} (\delta^\mu_\nu+ u^\mu_{\rm ZAMO} {u_{\rm ZAMO}}_\nu)S^\nu \,,
\end{align}
where $\hat{\Gamma} = -u_\mu u^\mu_{\rm ZAMO}$ is the relative boost factor. Now, one can project the relative spin vector to observer's local Cartesian triad with basis vectors
\begin{align}
e^{\hat{r}}_\mu &= (0,~\sqrt{g_{rr}},~0,~0) \,, \\
e^{\hat{\theta}}_\mu &= (0,~0,~\sqrt{g_{\theta\theta}},~0) \,, \\
e^{\hat{\phi}}_\mu &= (\frac{g_{t\phi}}{\sqrt{g_{\phi\phi}}},~0,~0,~\sqrt{g_{\phi\phi}}) \,,
\end{align}
to get the spin components with respect to this local orthonormal space triad
\begin{align}
\hat{S}^{\hat{i}} = \hat{S} (\cos\hat{\alpha}^s, \sin\hat{\alpha}^s\cos\hat{\beta}^s,\sin\hat{\alpha}^s\sin\hat{\beta}^s) \,.
\end{align}
The angles $\hat{\alpha}^s, ~\hat{\beta}^s$ represent the orientation of the spin. For a detailed description, please see Ref. 20.

The orbit configuration demonstrated in Fig. \ref{symspin}  represents the normal behavior of spinning particles, i.e., having equatorial symmetric patterns like the nonspinning cases. If one calculates the average value of z-coordinates when the particle passes through the x-z or y-z plane (i.e. the y= 0 plane or x = 0 plane), the average value will go to 0 after sufficient orbital evolution (see the bottom-right panel of Fig. \ref{symspin}). In this symmetric case, we get in x-z plane $\bar{z}_{y=0} = 6 \times 10^{-5}$ and in y-z plane $\bar{z}_{x=0} = 5 \times 10^{-5}$.  Another criterion is the difference of the maximum $|z|$ value achieved by the particle above and below the equatorial plane, i.e., $z_+ + z_-$.  For the symmetric pattern, $z_+$ equals $-z_-$ approximately, then $z_+ + z_- \approx 0$.

\begin{figure}
\begin{center}
\vspace{-20mm}
\includegraphics[height=3.0in]{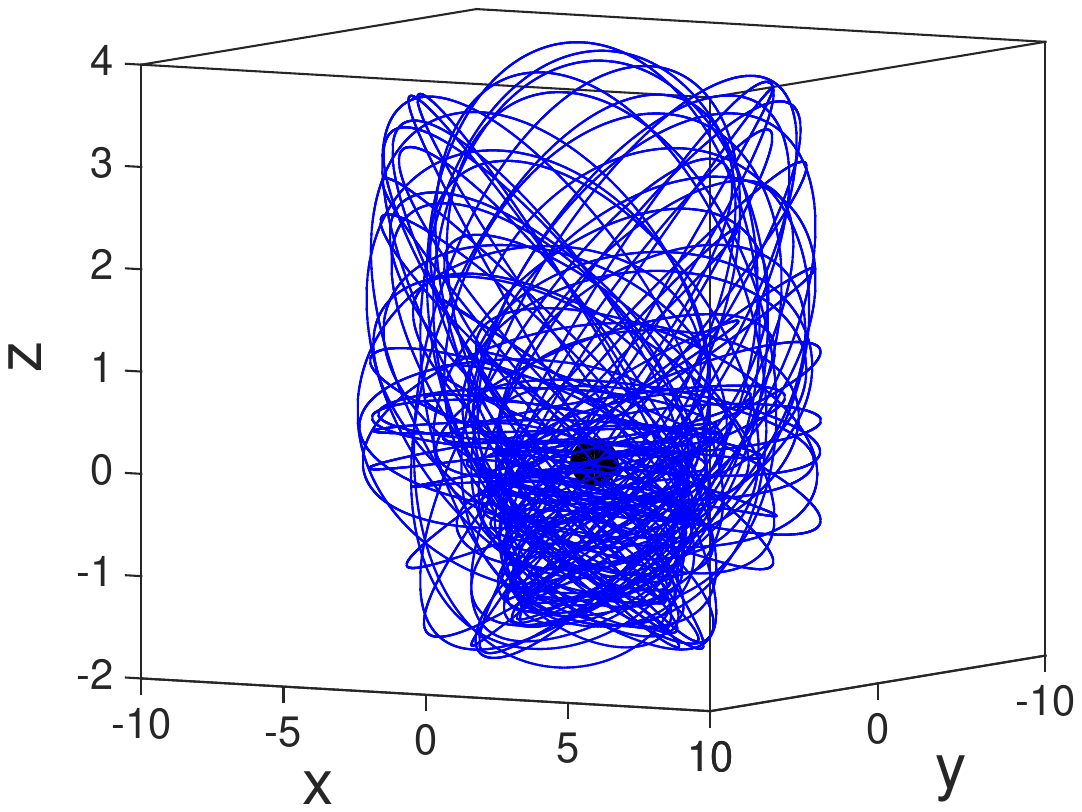}
\vspace{-30mm}
\includegraphics[height=3.0in]{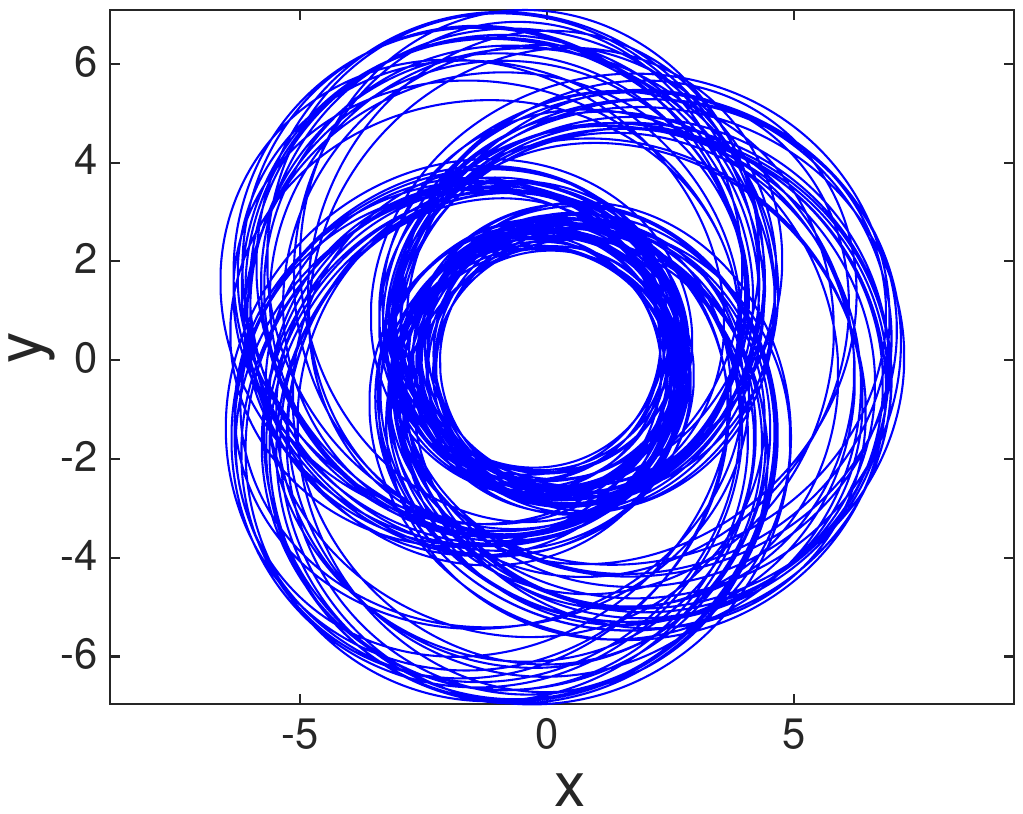}
\vspace{-10mm}
\includegraphics[height=3.0in]{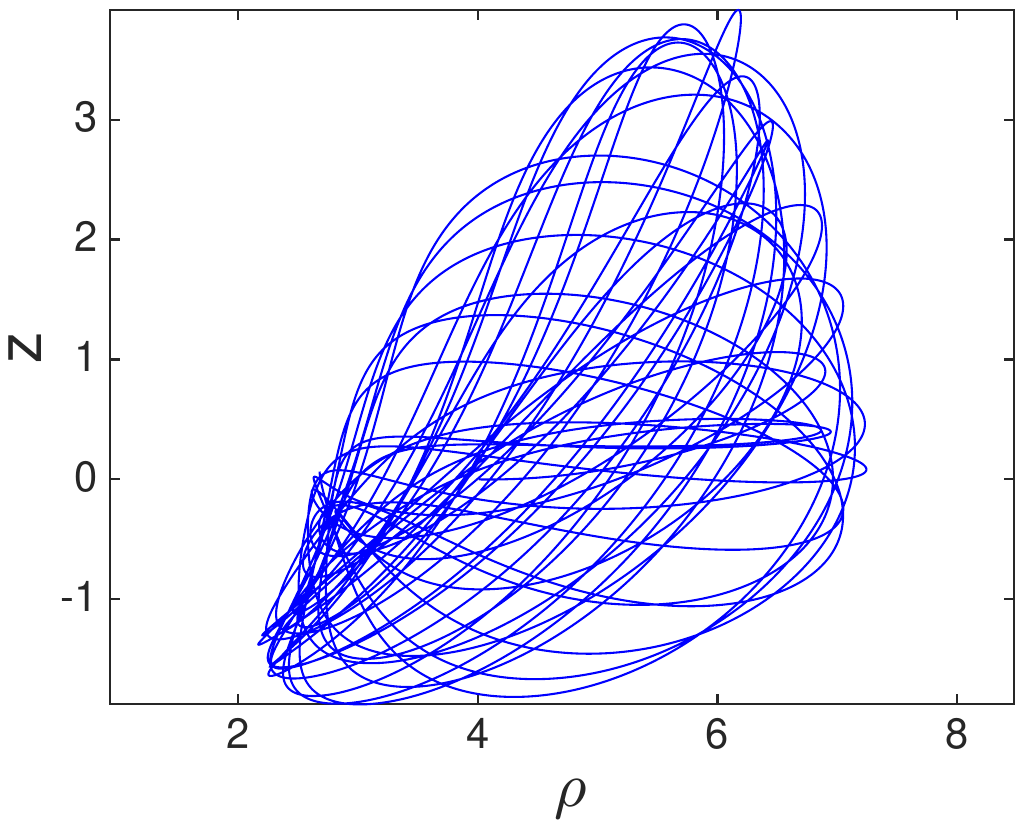}
\vspace{-10mm}
\includegraphics[height=3.0in]{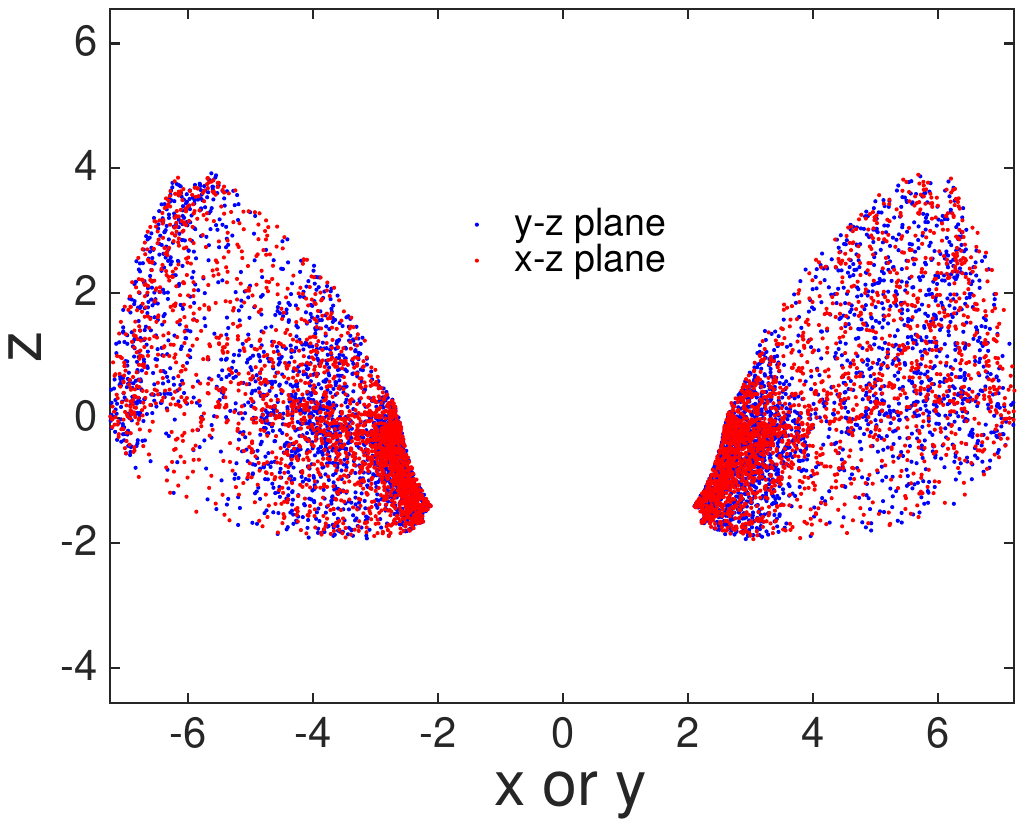}
\caption{Orbits of spinning particles with spin parameter $S = 0.8 $, energy $E = 0.9$ and total angular momentum $L_z = 2.5$ around a Kerr black hole with $a = 1$. These panels show a kind of ``upward" orbit, the particle is put at $r_0 = 4,~\theta_0=\pi/2, \phi_0 =0$ at beginning, and $u_0^\theta = 0$. The initial spin vector for the "upward" orbit is  $S_0^{r,~\theta} = (0.32, ~-0.08)$. The subsidiary data are 1.615, 0.187, 0.172, 0.594 and 0.192 for $u^t_0, ~u^r$, $u^\phi$, $S^t$ and $S^\phi$, respectively. The corresponding initial angles $\hat{\alpha}^s, ~\hat{\beta}^s$ are $61.9^\circ, ~293.6^\circ$. The top-left panel shows the 3D trajectories, the top-right and bottom-left ones show the projection orbits on $x-y$ and $\rho -z$ planes respectively, where $\rho =\sqrt{x^2+y^2}$. The bottom-right panel shows the projection points when the trajectories pass through the $y-z$ and $x-z$ planes.}  \label{asymspinu}
\end{center}
\end{figure}

\begin{figure}
\begin{center}
\vspace{-20mm}
\includegraphics[height=3.0in]{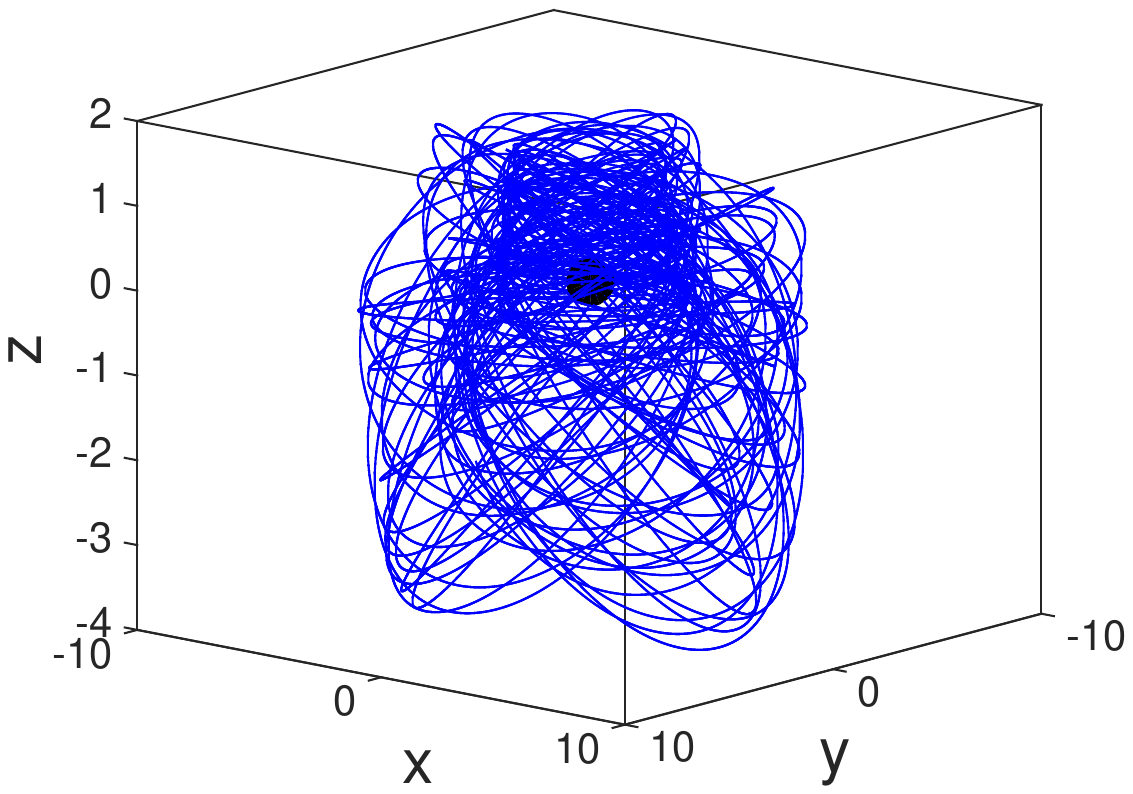}
\vspace{-30mm}
\includegraphics[height=3.0in]{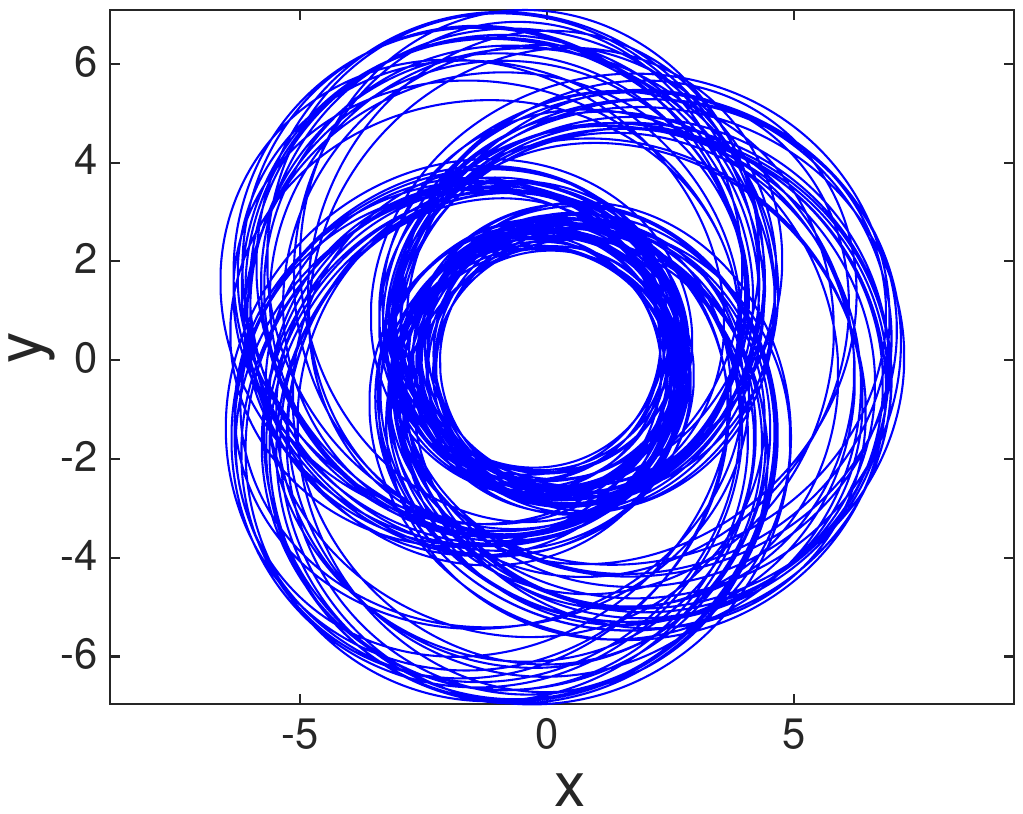}
\vspace{-10mm}
\includegraphics[height=3.0in]{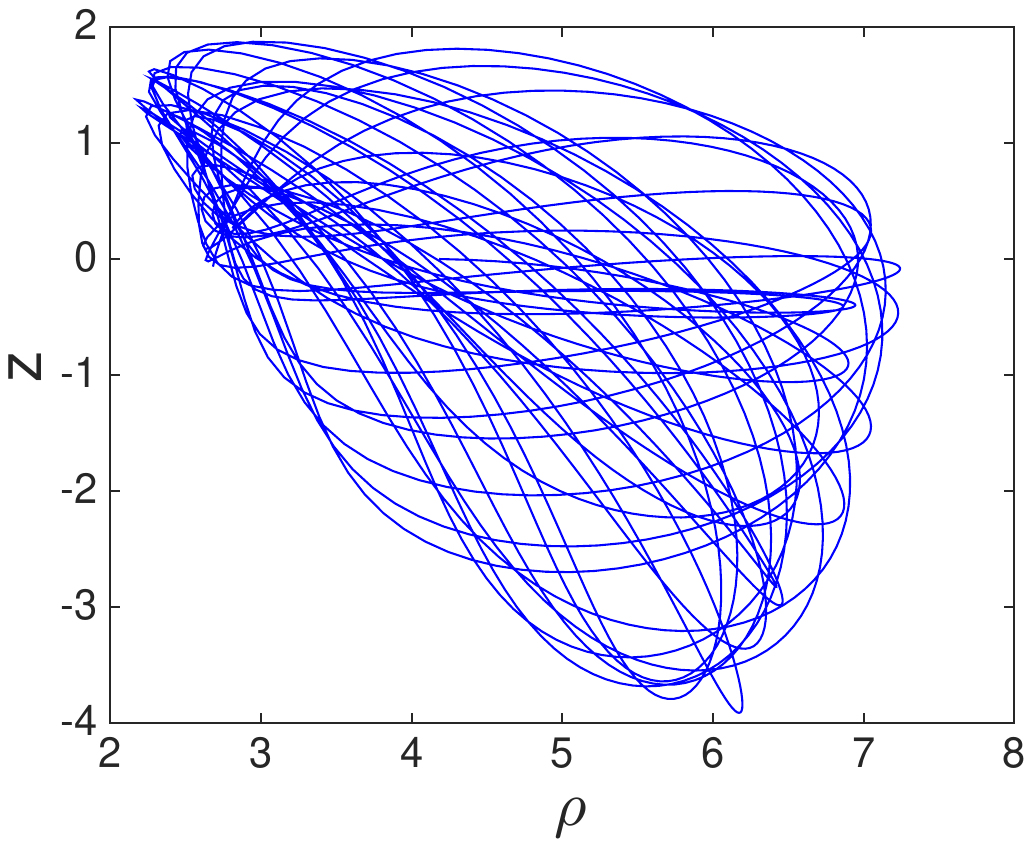}
\includegraphics[height=3.0in]{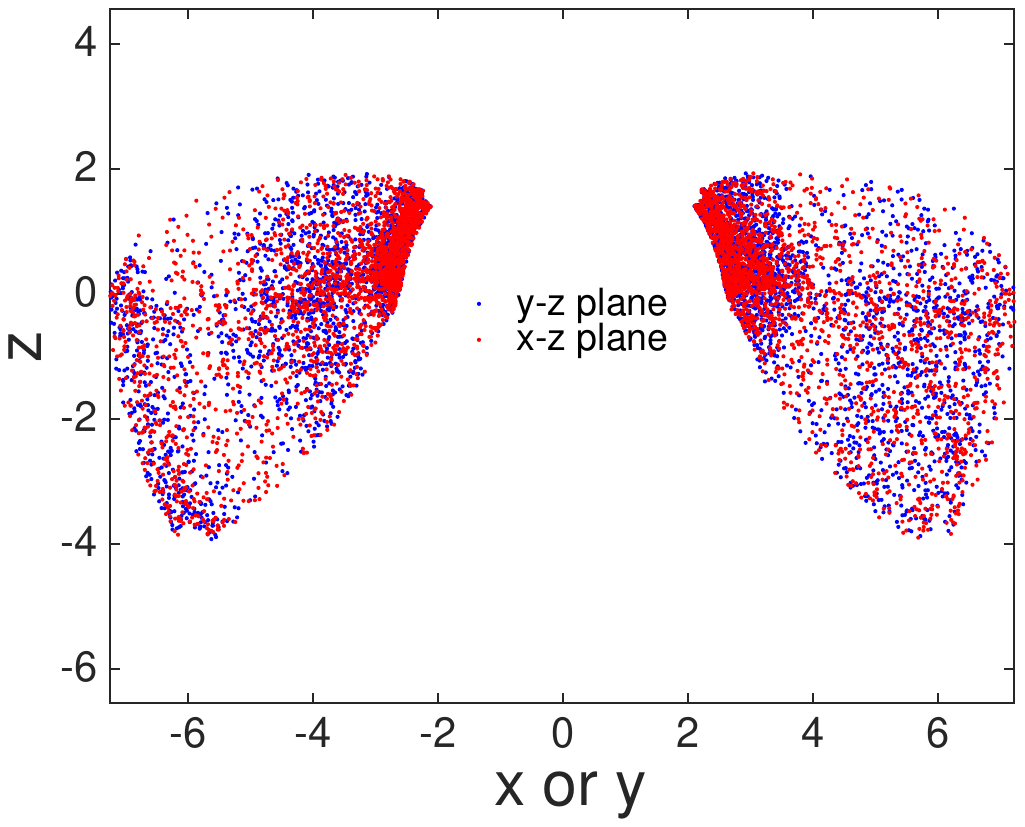}
\caption{Orbits of spinning particles with spin parameter $S = 0.8 $, energy $E = 0.9$ and total angular momentum $L_z = 2.5$ around a Kerr black hole with $a = 1$. These panels show a kind of ``downward" orbit, the particle is put at $r_0 = 4,~\theta_0=\pi/2, \phi_0 =0$ at beginning, and and $u_0^\theta = 0$. The initial spin vector for the "upward" orbit is  $S_0^{\mu} = (-0.32, ~-0.08)$. The subsidiary data are 1.615, -0.187, 0.172, -0.594 and -0.192 for $u^t_0, ~u^r$, $u^\phi$, $S^t$ and $S^\phi$, respectively. The corresponding initial angles $\hat{\alpha}^s, ~\hat{\beta}^s$ are $118.1^\circ, ~66.4^\circ$. The top-left panel shows the 3D trajectories, the top-right and bottom-left ones show the projection orbits on $x-y$ and $\rho -z$ planes respectively, where $\rho =\sqrt{x^2+y^2}$. The bottom-right panel shows the projection points when the trajectories pass through the $y-z$ and $x-z$ planes.}  \label{asymspind}
\end{center}
\end{figure}

However, from the simple analysis in Sec. 2, we know that the spin-spin coupling will supply a kind of ``force'' with different direction from the gravity of the mass. The direction of spin-spin interaction depends on the direction of spin. We also find that the spin-curvature coupling can destroy the reflection symmetry about the equatorial plane. For the generic orbits, the spin vector precesses along the trajectory in a very complicated way. In general, the spin directions are not reflectively symmetric about the equatorial plane, and then the total ``force'' is no longer symmetrical about the equatorial plane. In some situations, this asymmetry is enough obviously to be seen (as shown in Fig. \ref{asymspinu} and \ref{asymspind}). 

These orbits show a kind of exotic configuration, i.e., an asymmetry pattern appears about the equatorial plane of a Kerr black hole. It seems that the orbits have ``polarized'' directions. For example, we just change the initial velocity and the spin direction in the case of Fig. \ref{symspin}, and as a result we get an asymmetrical ``upward'' orbit (Fig. \ref{asymspinu}). In this case, the initial angles of the spin vector with respect to the local orthonormal space triad are $\hat{\alpha}^s_{\rm up}=61.9^\circ, ~\hat{\beta}^s_{\rm up}=293.6^\circ$. It means that spin points downward respect to the equatorial plane. Obviously, the pattern is asymmetric about the equatorial plane in the figure. From the bottom-left panel of Fig. \ref{asymspinu}, we can find that $z_+ + z_- \approx 2$, and the average $z$ values across y-z and x-z plane are 0.144 and 0.136 respectively (from the bottom-right panel). These two numbers deviate from 0 obviously comparing with the symmetric case. 

We keep all parameters except for the directions of momentum and spin ($\hat{\alpha}^s_{\rm down}=118.1^\circ, ~\hat{\beta}^s_{\rm down}=66.4^\circ$), then we get an asymmetrical ``downward polarized'' orbit in Fig. \ref{asymspind}. We notice that $\hat{\alpha}^s_{\rm up} + \hat{\alpha}^s_{\rm down} =180^\circ$ and $\hat{\beta}^s_{\rm up}+\hat{\beta}^s_{\rm down} =360^\circ$. It means that the initial direction of the spin in the downwards pattern points to upwards from the equatorial plane. Notice that this asymmetry is only about the equatorial plane, the orbital configuration is still axis-symmetric with sufficient evolution time. It looks like a force with one direction (along or anti-along with z axis) to push the particle floating above or sink down about the equatorial plane. This exotic asymmetrical phenomenon was  found in Ref. 18. In this paper, we study this phenomenon more thoroughly.


\begin{figure}
\begin{center}
\vspace{-20mm}
\includegraphics[height=3.0in]{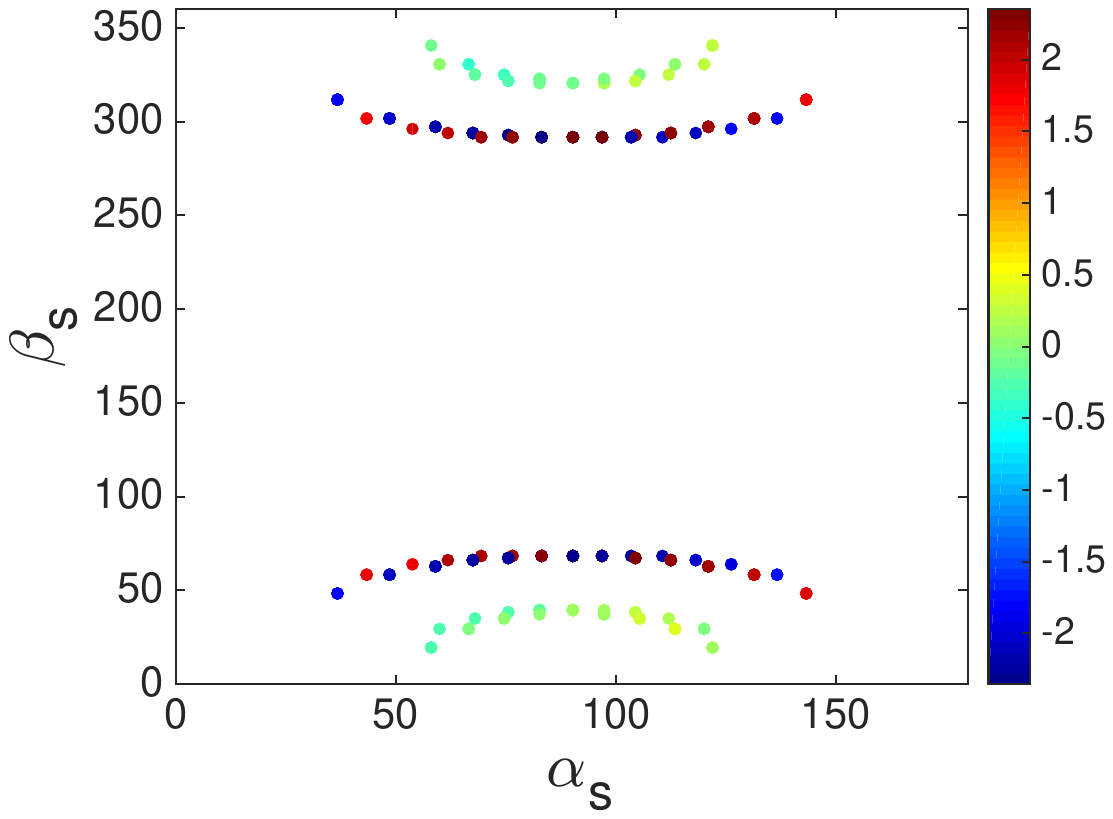}
\vspace{-30mm}
\includegraphics[height=3.0in]{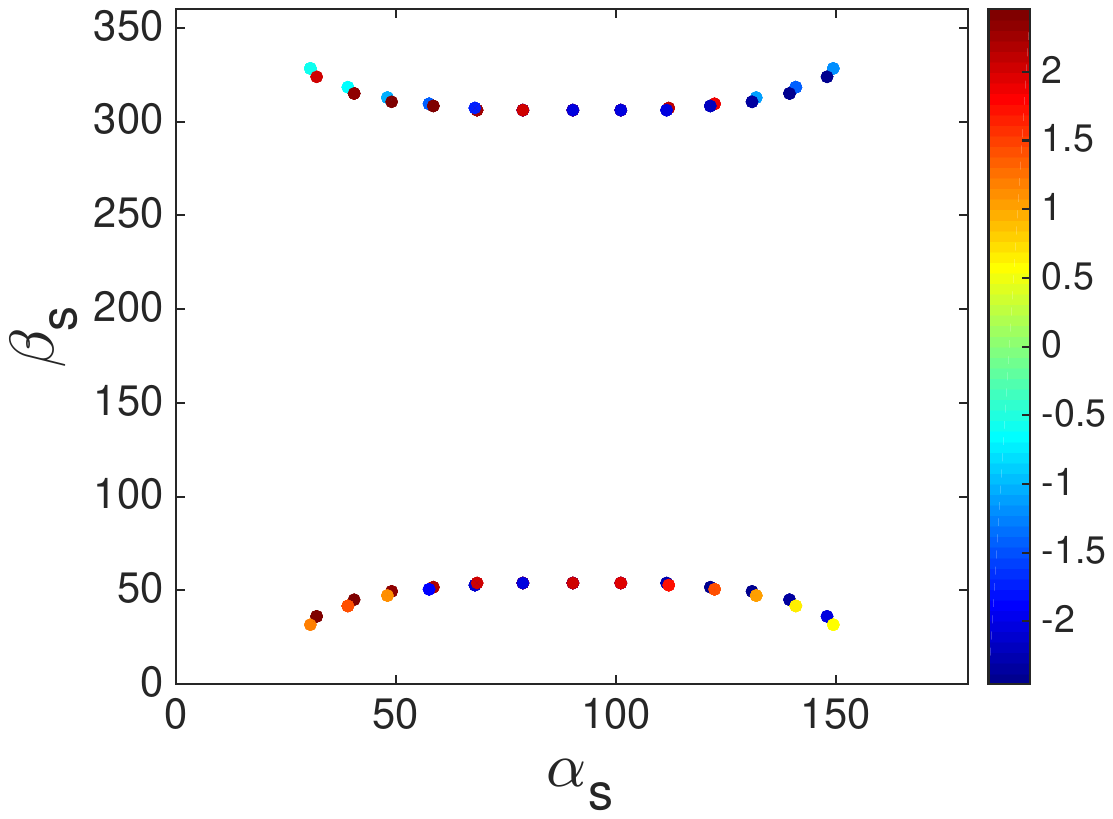}
\vspace{-10mm}
\includegraphics[height=3.0in]{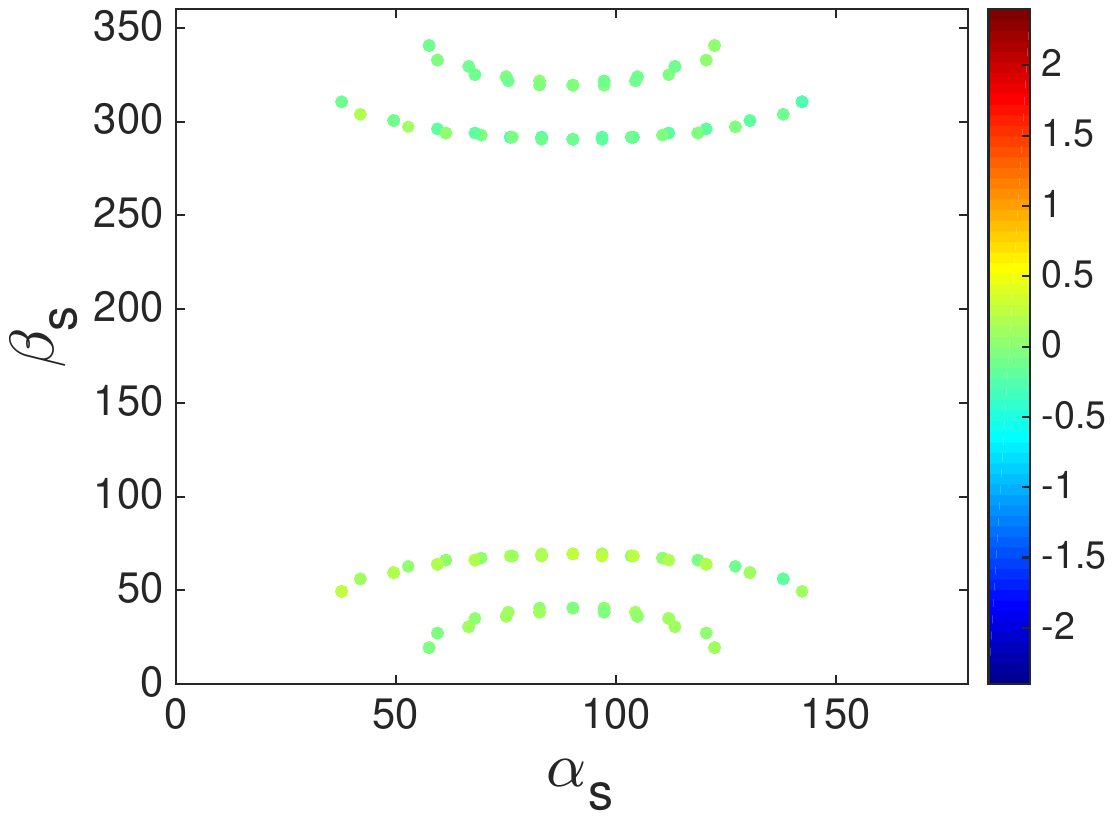}
\includegraphics[height=3.0in]{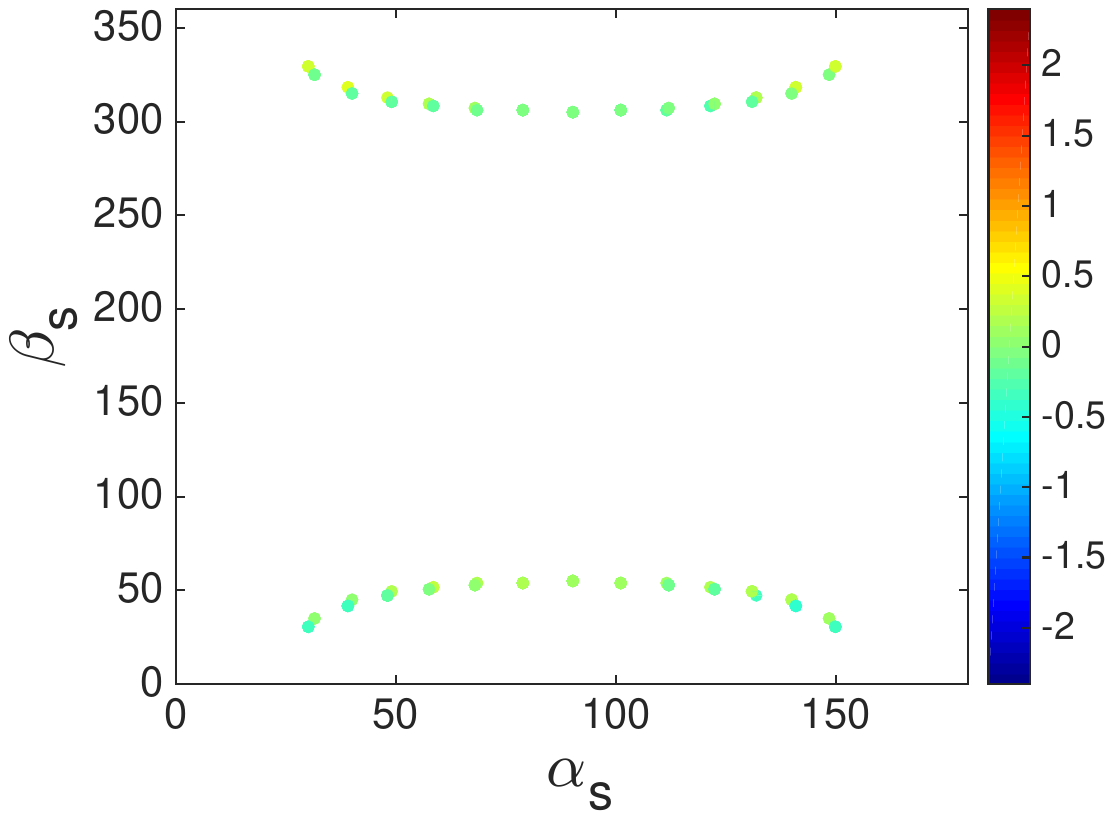}
\caption{(Color online) The values of $z_{+}+z_-$ with variable initial spin directions ($\hat{\alpha}^s, ~\hat{\beta}^s$). The color of point represents the values of $z_{+}+z_-$. Top-left: $E=0.9, ~L_z=2.5, ~a=1$ and $S=0.8$. Top-right: $E=0.9237, ~L_z=2.8, ~a=1$ and $S=0.8$. Bottom panels:  all parameters are the same but $S=0.4$.}  \label{dz}
\end{center}
\end{figure}

For revealing the relation between the orbital polarization orientation and the initial spin direction, in Fig. \ref{dz}, we plot the contour of $z^++z^-$ with variable angles $\hat{\alpha}^s, ~\hat{\beta}^s$. $z^+$ and $z^-$ mean the maximum and minimum of $z$ reached by a spinning particle with a set of given parameters after enough orbital evolution. We use varied color for different values of $z^++z^-$. Green points denote $z^++z^- \approx 0$, and dark red or blue ones denote $z^++z^-$ deviates from zero obviously. So red or blue points represent the asymmetry patterns. Obviously, $z^++z^- > 0$ implies an upwards orbit, and vice versa. It is clear that the orbital polarization direction is decided by the initial direction of the spin. Furthermore, we also give the results for a smaller spin value $S=0.4$, and do not find obvious asymmetric orbits (all points are green). 

Unfortunately we do not find a general quantitative criterion to determine which kind of initial conditions will produce asymmetric patterns. By a lot of scans in the parameter space, we can definitely conclude that the exotic orbits found by us can only happen with artificially large spin (i.e. $S \sim 1$). Actually, we did not find any obviously exotic orbits when $S=0.4$ for an extreme Kerr black hole. We can speculate carefully that the asymmetric phenomena cannot happen when $S < 0.1$. If we fix all parameters except for the spin components $S^r$ and $S^\theta$, which are equivalent to the angles $\alpha^s$ and $\beta^s$, we may determine the range of the angles that the exotic behavior occurs. From the top-left panel of Fig. \ref{dz}, one can find that when $|\beta^s| > 50^\circ $ (e.g. we can take $310^\circ$ as the same as $-50^\circ$) the asymmetric behavior happens.  For the cases of initial angle $|\beta^s| < 50^\circ $, patterns of orbits are approximately symmetric. Be careful, this criterion is only correct for the special case of $E=0.9, ~L_z=2.5, ~a=1$ and $S=0.8$. If changing any one of these four parameters, the range of angles for exotic behaviors is different.

We emphasize that there is no direct connection of the nonreflection symmetric orbits with chaotic behaviors. Some nonreflection symmetric orbits can be regular (for example a case with energy 0.9237, angular momentum 2.8 and spin magnitude 0.8), and some may be chaotic. Such kind of exotic phenomenon is restored for very large evolution time (we evolve the orbits up to $10^7$ M), we believe that it is not a transient phenomenon. Notice that in Figs. \ref{asymspinu}-\ref{dz}, we choose the extreme Kerr black hole just for demonstrating the most prominent effects. However, there is no connection between the asymmetry and the naked singularity ($a = 1$). An immediate example is in the case of $ E= 0.9237, L_z = 2.8$ and $S= 0.8$, instead of $a=1$ with $a=0.998$, the asymmetric pattern happens too.

Additionally, we do not find any obviously asymmetric pattern if the black hole is a Schwarzschild one in the parameter space we scanned (energy from 0.8 to 0.95, angular momentum from 2.0 to 5.0 and $r_0$ from 8 to 10). This may imply that only the spin-spin interaction but not spin-orbit one causes the exotic phenomena. However, our scanning do not cover all the parameter space. This statement may be only valid for the parameter space we scanned. 

For the Kerr spacetime, the spin-spin coupling is a necessary condition but not a sufficient one for the appearance of asymmetric patterns. Without this spin-spin effect between the spinning particle and fast rotating black hole, asymmetric patterns may not happen. However, the spin-spin interaction does not guarantee the appearance of asymmetric orbits. A fast rotating Kerr black hole can be easily found in the universe, but the spin magnitude of particle should be treated carefully. As we discussed in the Sec. 2, the physical value of spin must be $\ll 1$ for an extreme-mass-ratio system. That means it is difficult to find such exotic orbits in the realistic astrophysics. Therefore, our finding may have no influence on the gravitational-wave detection of LISA, Taiji, and Tianqin.

\section{Conclusions}
It is a well-known fact that the gravitational force is an attractive force. However, as already revealed by a few researchers, we know that the spin-spin interaction between the spinning particle and Kerr black hole can have arbitrary action directions (e.g. Ref. 31). Phenomenologically, the spin-spin coupling can actually offer a kind of ``counter-gravity''. The exotically asymmetrical orbit configurations about the equatorial plane demonstrated in Figs. \ref{asymspinu} and \ref{asymspind} should come from the spin-spin coupling, because we have not found this asymmetry either for nonspinning particles or for the Schwarzschild black hole. However, the orbits in Figs. \ref{asymspinu} and \ref{asymspind} are quite complicated, and in this paper we do not plan to analyze the quantitative relation between the asymmetry and spin-spin interaction.  As analyzed in Sec. 2, the existence of spin can destroy the reflection symmetry about the equatorial plane. This may give a clue to the physical origin of the asymmetrical phenomena. 
 
However, not all the spinning particles demonstrate such asymmetrical orbit patterns,  the asymmetry appears only for cases with special physical parameters. We still have not found a criterion to determine if a spinning particle with a certain set of parameters will have an asymmetrical orbit shape, but the numerical results show that it may easier to appear for large eccentricities. Actually, we do not give a critical spin value for the appearance of asymmetry because it depends on too many parameters. There is, however, no evidence that asymmetrical phenomena happen when dimensionless spin magnitude $S \ll 1$.  We conclude that the asymmetry can only happen for astrophysically irrelevant large spin values. On the other hand, it is  interesting to study the complicated behavior and dynamical nature of these extreme spinning particles.

For comparable mass-ratio binary systems, the spin of both components can be\\$\sim1$. Until now, there is no report on the analogous asymmetrical orbits for comparable mass-ratio binaries. It is very interesting to investigate if there are asymmetrical orbits in the comparable mass-ratio binary systems or not. The phenomena revealed in this paper should be interesting for the study of dynamical properties of the spinning particles in strong gravitational field. The gravitational waves from the asymmetrical orbits should have some obvious properties which distinguish from the normal orbits. However, considering the asymmetry can only appear in the astrophysically unrealistic cases, it should have no influence on the gravitational wave detections. More detailed studies on this asymmetry should be done in the future works.


\section*{Acknowledgements}
We appreciate the anonymous Referee for pointing an error in Eq. (\ref{eleven}). This work is supported by NSFC No. U1431120, QYZDB-SSW-SYS016 of CAS; W.-B. Han is also supported by Youth Innovation Promotion Association CAS.
\\


\end{document}